\documentstyle[12pt]{article}

\def\beq{\begin{equation}} \def\eeq{\end{equation}}
\def\bea{\begin{eqnarray}} \def\eea{\end{eqnarray}}
\let\nn=\nonumber
\def\beann{\begin{eqnarray*}} \def\eeann{\end{eqnarray*}}

\newcommand{\append}[1]{\protect\stepcounter{section}
		        \setcounter{equation}{0} 
		        \section*{Appendix \thesection \, #1}
		        \addcontentsline{toc}{section}{Appendix
		        \thesection: #1}}

\let\a=\alpha \let\be=\beta \let\g=\gamma \let\de=\delta
\let\e=\varepsilon  \let\h=\eta \let\th=\theta
\let\dh=\vartheta   \let\m=\mu
  \let\p=\pi  \let\s=\sigma
 \let\ps=\psi
  \let\PH=\Phi 
\let\Om=\Omega  
\let\La=\Lambda  

\let\qd=\quad  

\def\tst#1{{\textstyle #1}}
\def\dst#1{{\displaystyle #1}}

\def\0{\over } \def\1{\vec }     \def\2{{1\over2}} \def\4{{1\over4}}
\def\5{\bar }  \def\6{\partial } \def\7#1{{#1}\llap{/}}

\def\<{\langle } \def\>{\rangle }

\let\auf=\uparrow \let\ab=\downarrow

  \def\CO{{\cal O}}
\def\i{{\rm i}} \def\tr{\mbox{tr}}

\renewcommand{\det}{\mbox{det}}
\renewcommand{\Im}{\mbox{Im}}
\renewcommand{\Re}{\mbox{Re}}

\def\sign{\mbox{sign}}  

\def\ssign{\mbox{\scriptsize sign}}

\begin{document}
\thispagestyle{empty}
\begin{center}
{\Large {\bf Time and temperature dependent correlation functions of
the 1D impenetrable electron gas\\}}
\vspace{7mm}
{\large F.~G\"{o}hmann$^\dagger$\footnote
{e-mail: goehmann@insti.physics.sunysb.edu},
A.~G.~Izergin$^\ddagger$\footnote
{e-mail: izergin@pdmi.ras.ru},
V.~E.~Korepin$^\dagger$\footnote
{e-mail: korepin@insti.physics.sunysb.edu},
A.~G.~Pronko$^\ddagger$\footnote
{e-mail: agp@pdmi.ras.ru}}\\
\vspace{5mm}
$^\dagger$Institute for Theoretical Physics,\\ State University of New
York at Stony Brook,\\ Stony Brook, NY 11794-3840, USA\\
$^\ddagger$Sankt-Petersburg Department of the V.\ A.\ Steklov
Mathematical Institute,\\
Fontanka 27, 191 011 Sankt-Petersburg, Russia
\vspace{20mm}

{\large {\bf Abstract}}
\end{center}
\begin{list}{}{\addtolength{\rightmargin}{10mm}
               \addtolength{\topsep}{-5mm}}
\item
We consider the one-dimensional delta-interacting electron gas in the
case of infinite repulsion. We use determinant representations to
study the long time, large distance asymptotics of correlation
functions of local fields in the gas phase. We derive differential
equations which drive the correlation functions. Using a related
Riemann-Hilbert problem we obtain formulae for the asymptotics of the
correlation functions, which are valid at all finite temperatures.
At low temperatures these formulae lead to explicit asymptotic
expressions for the correlation functions, which describe power law
behavior and exponential decay as functions of temperature, magnetic
field and chemical potential.\\[2ex]
{\it PACS:} 05.30.Fk; 71.10.Pm; 71.27.+a\\
{\it Keywords: strongly correlated electrons; temperature correlations;
quantum correlation functions; determinant representation}
\end{list}

\clearpage

\section{Introduction}
We consider the electron gas with delta interaction in one space and
one time dimension. It can be described by canonical Fermi fields
$\ps_\a (x)$ with canonical equal time anticommutation relations
\beq
     \ps_\a (x) \ps_\be^\dagger (y) + \ps_\be^\dagger (y) \ps_\a (x)
        = \de^\a_\be \de (x - y) \qd.
\eeq
The spin index $\a$ runs through two values, $\a = \auf, \ab$. The
Hamiltonian of the model is
\beq \label{ham}
     H = \int_{- \infty}^\infty dx \left\{
         \6_x \ps_\a^\dagger \6_x \ps_\a
	 + c : \! \left( \ps_\a^\dagger \ps_\a \right)^2 \! :
	 - \, \m \, \ps_\a^\dagger \ps_\a + B (\ps^\dagger \s^z \ps)
	 \right\} \qd.
\eeq
Here $\m$ is the chemical potential, and $B$ is the magnetic field.
$\s^z$ is a Pauli matrix, and $c$ is the coupling constant. In this
paper we shall consider the case of infinite repulsion $c = \infty$.

The model can be solved exactly \cite{Yang67}. Its thermodynamics
was described in \cite{Takahashi71}. The pressure
\beq \label{press}
     P = \frac{T}{2\p} \int_{- \infty}^\infty dk
         \ln \left( 1 + e^{(\m + B - k^2)/T} + e^{(\m - B - k^2)/T}
	     \right)
\eeq
serves as thermodynamic potential. $T$ is the temperature. Note that
the expression for $P$, eq.\ (\ref{press}), is formally the same as
for free fermions with effective chemical potential $\m +
T \ln(2 \cosh(B/T))$. Thus, there are two different zero temperature
phases depending on whether $\lim_{T \rightarrow 0+} (\m +
T \ln(2 \cosh(B/T)) = \m + |B|$ is positive or negative. For $\m + |B|
> 0$ the density $D = \6 P/\6 \m$ has a positive limit as $T$ goes to
zero. For $\m + |B| < 0$ the density at zero temperature vanishes. This
is the phase we are interested in. We call it the gas phase. We shall
assume in the following that $\m + T \ln(2 \cosh(B/T)) < 0$.

The number of up-spin particles and the number of down-spin particles
are separately conserved. The densities $D_\auf$ of up-spin electrons
and $D_\ab$ of down-spin electrons are obtained as
\bea \label{densauf}
     D_\auf & = & \frac{\6 P}{\6(\m - B)} \: = \: \frac{1}{2\p}
                  \int_{- \infty}^\infty dk \; \frac{e^{- B/T}}
		  {2 \cosh(B/T) + e^{(k^2 - \m)/T}} \qd, \\
		  \label{densab}
     D_\ab & = & \frac{\6 P}{\6(\m + B)} \: = \: \frac{1}{2\p}
                  \int_{- \infty}^\infty dk \; \frac{e^{B/T}}
		  {2 \cosh(B/T) + e^{(k^2 - \m)/T}} \qd.
\eea
It is obvious from these expressions that they satisfy the
relation
\beq
     D_\ab (B) = D_\auf (-B) \qd.
\eeq
The density $D$ of the gas is
\beq \label{density}
     D = \frac{\6 P}{\6 \m} = D_\auf + D_\ab \qd.
\eeq
Note that the density vanishes exponentially at small temperatures,
\beq \label{lowtdens}
     D \; = \; \left\{ \begin{array}{l}
               \sqrt{\frac{T}{\p}} \, e^{\m/T} \\[1ex]
               \frac{1}{2} \sqrt{\frac{T}{\p}} \, e^{(\m + |B|)/T}
	       \end{array} \right.
	       \qd \mbox{if} \qd
	       \begin{array}{l} B = 0 \qd, \\[1ex] B \ne 0 \qd.
	       \end{array}
\eeq
At very low temperatures density and pressure are related by the
ideal gas law,
\beq
     P = DT \qd.
\eeq
For larger temperatures there are corrections to the ideal gas law.

In this article we study the two-point correlation functions
$G_\a^\pm (x,t)$, $\a = \auf$, $\ab$, of local fields. We shall follow
the approach of the book \cite{KBIBo} to correlation functions of
solvable models. The definitions of $G_\a^\pm (x,t)$ are given in the
next section, where we recall a determinant representation for
$G_\a^\pm (x,t)$, \cite{IzPr97,IzPr98}, which is the starting point
of our considerations. The correlation functions $G_\a^\pm (x,t)$ are
not elementary functions. They are determined by a pair of coupled
nonlinear partial differential equations \cite{IIKS90} which is derived
in section 3. In section 4 we shall formulate a corresponding
Riemann-Hilbert problem \cite{IIKV92,ZaSh79,FaTa87}. It will fix the
solution of the differential equations, which is relevant for the
calculation of the correlation functions. Furthermore, the asymptotic
solution of the Riemann-Hilbert problem for large $x$ and $t$ ($x/t$
fixed) can be obtained by reducing it to a known form
\cite{Manakov74,ZaMa76,Its81,ItNo86,DIZ93,DeZh93}. This is done in
section 5. Using the asymptotic solution of the Riemann-Hilbert
problem and the differential equations we shall obtain the asymptotics
of the correlation functions $G_\a^\pm (x,t)$ in terms of a contour
integral in section 6. In section 7 we shall present explicit asymptotic
expressions for $G_\a^\pm (x,t)$ at low temperatures. The asymptotic
expressions for the correlation functions $G_\a^\pm (x,t)$ at low
temperatures are products of correlation functions of free fermions
and corrections caused by the interaction. The leading factors of
the correction terms describe exponential decay. The next to leading
factors are power law corrections. The leading factor of the asymptotics
of the spin-up correlation function $G_\auf^+ (x,t)$, for instance, is
\beq \label{appetizer}
     G_\auf^+ (x,t) = e^{\i t(\m - B)} e^{\i x^2/4t} e^{- x D_\ab} \qd.
\eeq
The factor $e^{\i t(\m - B)} e^{\i x^2/4t}$ is the exponential part of
the free fermionic correlation function. The
factor $e^{- x D_\ab}$ appears due to the interaction. The explicit
form of the power law correction as a function of temperature, chemical
potential and magnetic field will be given in section 7 (see eqs.\
(\ref{gpas}), (\ref{nugamma}), (\ref{gpas2}), (\ref{nuzp})). The
occurrence of the density of down-spin electrons in equation
(\ref{appetizer}) has a clear physical interpretation. Due to the Pauli
principle and the locality of the interaction, up-spin electrons are
scattered only by down-spin electrons. $D_\ab$ appears as the
reciprocal of the correlation length in eq.\ (\ref{appetizer}). The
higher the density of down-spin electrons, the more often an up-spin
electron is scattered and the smaller is the correlation length.
\section{A determinant representation for correlation functions}
We shall consider the following two-point correlation functions of
local fields,
\bea \label{gp1}
     G_\a^+ (x,t) & = & \frac{\tr \left( e^{- H/T} \,
        \ps_\a (x,t) \ps_\a^\dagger (0,0) \right)}
	{\tr \left( e^{- H/T} \right)} \qd, \\ \label{gm1}
     G_\a^- (x,t) & = & \frac{\tr \left( e^{- H/T} \,
        \ps_\a^\dagger (x,t) \ps_\a (0,0) \right)}
	{\tr \left( e^{- H/T} \right)} \qd,
\eea
where $\a = \auf, \ab$. These correlation functions depend not only
on space and time variables $x$ and $t$, but also on the temperature
$T$, the chemical potential $\m$ (or the density $D$) and the magnetic
field $B$. Due to the invariance of the Hamiltonian (\ref{ham}) under
the transformation $\auf \: \rightleftharpoons \, \ab$,
$B \rightleftharpoons - B$, we have the identity
\beq \label{aufab}
     G_\ab^\pm (x,t) = \left. G_\auf^\pm (x,t)
                       \right|_{B \rightleftharpoons - B} \qd.
\eeq
Hence, we can restrict our attention in the following to the correlation
functions $G_\auf^\pm (x,t)$. These correlation functions were
represented by means of determinants of integral operators
\cite{IzPr97,IzPr98}. Now we shall recall these formulae:
\bea \label{gp2}
     G_\auf^+ (x,t) & = & \frac{- e^{\i t (\m - B)}}{2 \p}
        \int_{- \p}^\p \! d\h \, \frac{F(\g,\h)}{1 - \cos(\h)} \,
	b_{++} \det\left( \hat I + \g \hat V \right) \qd, \\
     \label{gm2}
     G_\auf^- (x,t) & = & \frac{e^{- \i t (\m - B)}}{4 \p \g}
        \int_{- \p}^\p \! d\h \, F(\g,\h) B_{--}
	\det\left( \hat I + \g \hat V \right) \qd.
\eea
It will take us some time to define notations. First of all
\beq
     F(\g,\h) = 1 + \frac{e^{\i \h}}{\g - e^{\i \h}} +
                    \frac{e^{- \i \h}}{\g - e^{- \i \h}}
\eeq
and
\beq
     \g = 1 + e^{2B/T} \qd.
\eeq
$\hat V$ is an integral operator with kernel $V(k,p)$. It can be
represented in the ``standard form'' (see page 316 of the book
\cite{KBIBo})
\beq \label{kerv}
     \g V(k,p) = \frac{e_+(k) e_-(p) - e_+(p) e_-(k)}{k - p} \qd.
\eeq
Here
\bea
     e_-(k) & = & \sqrt{\frac{\g \dh(k)}{\p}} e^{\tau(k)/2} \qd, \\
     \label{ep}
     e_+(k) & = & \frac{1}{2} \sqrt{\frac{\g \dh(k)}{\p}} e^{-\tau(k)/2}
                  \left\{ (1 - \cos(\h)) e^{\tau(k)} E(k) + \sin(\h)
		  \right\} \qd, \\
     \g \dh(k) & = & \frac{2 \cosh(B/T)}{2 \cosh(B/T) +
                     e^{(k^2 - \m)/T}} \qd, \\
     \tau(k) & = & \i k^2 t - \i k x
\eea
and
\beq
     E(k) = \mbox{p.v.} \int_{- \infty}^\infty \! dp \,
            \frac{e^{- \tau(p)}}{\p (p - k)} \qd.
\eeq
The integral operator (\ref{kerv}) closely reminds an integral operator
which appears in the theory of correlation functions of the
impenetrable Bose gas (cf.\ page 346 of \cite{KBIBo}). The inverse
of $\left( \hat I + \g \hat V \right)$ can be defined in the following
way,
\beq
     \left( \hat I + \g \hat V \right) \left( \hat I - \g \hat R \right)
        = \hat I \qd.
\eeq
The resolvent $\hat R$ is an integral operator with kernel $R(k,p)$.
It can be represented in the ``standard form'',
\beq \label{kerr}
     \g R(k,p) = \frac{f_+(k) f_-(p) - f_+(p) f_-(k)}{k - p} \qd,
\eeq
where the functions $f_\pm(k)$ can be defined as solutions of the
integral equations
\beq \label{fies}
     f_\pm(k) + \int_{-\infty}^\infty \! dp \, \g V(k,p) f_\pm(p)
        = e_\pm (k) \qd.
\eeq
This theorem is proven on page 318 of the book \cite{KBIBo}. We should
mention that both kernels $V(k,p)$ and $R(k,p)$ are symmetric,
\beq
     V(k,p) = V(p,k) \qd, \qd R(k,p) = R(p,k) \qd.
\eeq
At this point it is convenient to define potentials $B_{ab}$ and
$C_{ab}$,
\beq \label{defpot}
     B_{ab} = \dst{\int_{- \infty}^\infty \! dk \, e_a(k) f_b(k)}
        \qd, \qd
     C_{ab} = \dst{\int_{- \infty}^\infty \! dk \, k e_a(k) f_b(k)}
        \qd.
\eeq
Here both indices $a$ and $b$ run through two values, $a, b = \pm$.
{For} example, the factor $B_{--}$ in (\ref{gm2}) is defined as
\beq
     B_{--} = \int_{- \infty}^\infty \! dk \, e_-(k) f_-(k) \qd.
\eeq
In order to define $b_{++}$ we need the function
\beq
     G(x,t) = \frac{1}{2\p} \int_{- \infty}^\infty \! dk \,
              e^{- \tau(k)} = \frac{e^{- \i \p /4}}{2 \sqrt{\p t}}
	      e^{\i x^2/4t} \qd.
\eeq
The factor $b_{++}$ in (\ref{gp2}) is defined as
\beq
     b_{++} = B_{++} - (1 - \cos(\h)) G(x,t) =
              \int_{- \infty}^\infty \! dk \, e_+(k) f_+(k)
	      - (1 - \cos(\h)) G(x,t) \qd.
\eeq
In such a way we accomplished the description of the determinant
representation of the correlation functions (\ref{gp2}), (\ref{gm2}). It
reminds the determinant representation for correlation functions of
the impenetrable Bose gas (see page 346 of the book \cite{KBIBo}).
Let us also mention that the symmetry of $V(k,p)$ leads to $B_{+-} =
B_{-+}$.

Let us discuss the factor $1 - \cos(\h)$ in the denominator of
(\ref{gp2}). Let us show that there is no singularity at $\h = 0$.
Indeed, $e_+(k)$, eq.\ (\ref{ep}), is proportional to $\h$ as
$\h \rightarrow 0$. Hence $b_{++} \sim \h^2$, and there is no
singularity. Let us change the integration variable to $z = e^{\i \h}$.
Then
\bea \label{gp3}
     G_\auf^+ (x,t) & = & \frac{e^{\i t (\m - B)}}{\p \i}
        \oint \! dz \, \frac{F(z)}{(z - 1)^2} \,
	b_{++}(z) \det\left( \hat I + \g \hat V \right)(z) \qd, \\
     \label{gm3}
     G_\auf^- (x,t) & = & \frac{e^{- \i t (\m - B)}}{4 \p \i \g}
        \oint \! dz \, \frac{F(z)}{z} B_{--}(z)
	\det\left( \hat I + \g \hat V \right)(z) \qd.
\eea
Here the contour of integration is the unit circle, and
\beq
     F(z) = 1 + \frac{z}{\g - z} + \frac{1}{\g z - 1} \qd.
\eeq
\section{Differential equations}
Starting from a determinant representation for a quantum correlation
function it is possible to derive differential equations which
drive this correlation function (see \cite{IIKS90} and chapter XIV
of the book \cite{KBIBo}). In our specific case calculations similar
to the ones presented on pages 345 - 353 of the book \cite{KBIBo} lead
us to the following differential equations,
\beq \label{deqs}
     \begin{array}{rcr}
      - \; \i \, \dst{\frac{\6 b_{++}}{\6 t}} & = &
           \dst{\frac{\6^2 b_{++}}{\6 x^2}}
           + 2 b_{++}^2 B_{--} \qd, \\[2ex]
        \i \, \dst{\frac{\6 B_{--}}{\6 t}} & = &
	   \dst{\frac{\6^2 B_{--}}{\6 x^2}}
           + 2 b_{++} B_{--}^2 \qd.
     \end{array}
\eeq
We should notice that a change of variables,
\beq
     x = - 2 \tilde x \qd, \qd t = 2 \tilde t \qd,
\eeq
leads to coincidence of our system of eqs.\ (\ref{deqs}) with the
system (5.47) on page 344 of the book. The system of equations
(\ref{deqs}) is called separated nonlinear Schr\"odinger equation.
It is completely integrable. It has many solutions. The modern way to
fix a solution of the system (\ref{deqs}) is to formulate a
corresponding Riemann-Hilbert problem (see chapter XV of the book
\cite{KBIBo}). We shall formulate this Riemann-Hilbert problem in the
next section. We will be able to solve the Riemann-Hilbert problem
asymptotically as $x \rightarrow \infty$ and $t \rightarrow \infty$
($x/t$ fixed). This will give us asymptotic expressions for $b_{++}$
and $B_{--}$. The solution of the Riemann-Hilbert problem will also
give us asymptotic expressions for the other potentials $B_{ab}$ and
$C_{ab}$ (\ref{defpot}).

Let us now explain how to express $\det(\hat I + \g \hat V)$ in terms
of these potentials. The logarithmic derivatives of the determinant
can be expressed in terms of $B$ and $C$,
\beq \label{dxtdet}
     \begin{array}{rcl}
        \6_x \, \ln \det \left( \hat I + \g \hat V \right)
	   & = & \i B_{+-} = \i B_{-+} \qd, \\[2ex]
        \6_t \, \ln \det \left( \hat I + \g \hat V \right)
	   & = & - \i (C_{+-} +  C_{-+} + (1 - \cos(\h))G B_{--}) \qd.
     \end{array}
\eeq
We shall also mention several identities, which will be useful for
further calculations:
\bea \label{dxbpm}
     \6_x B_{+-} & = & - \i B_{--} b_{++} \qd, \\
     C_{--} & = & \i \6_x B_{--} - B_{--} B_{+-} \qd, \\
     C_{++} & = & - \i \6_x B_{++} - 2 (1 - \cos(\h)) G B_{+-} +
                  B_{+-} B_{++} \qd, \\
     C_{+-} - C_{-+} & = & B_{+-}^2 - B_{++} B_{--} \qd.
\eea
In the next section we shall introduce a Riemann-Hilbert problem and
explain how one can express the potentials $B_{ab}$ and $C_{ab}$ in
terms of its solution $\chi(k)$.
\section{Riemann-Hilbert problem}
The modern way to solve completely integrable differential equations
is by means of an equivalent Riemann-Hilbert problem
\cite{IIKV92,ZaSh79,FaTa87}. Solving a Rie\-mann-Hilbert problem amounts
to constructing a piecewise analytic matrix-func\-tion. Let us formulate
the Riemann-Hilbert problem connected to the separated nonlinear
Schr\"odinger equation (\ref{deqs}). Let us introduce a
$2 \times 2$-matrix $\chi (k)$.  The matrix $\chi (k)$ should be
analytic as a function of $k$ in the upper half plane, $\Im (k) > 0$,
and in the lower half plane, $\Im (k) < 0$. In general, it has
different limits $\chi_+ (k)$ as $k$ approaches the real axis from
above and $\chi_- (k)$ as $k$ approaches the real axis from below.
These limits are related by means of a conjugation matrix $G(k)$,
\beq
     \chi_- (k) = \chi_+ (k) G(k) \qd \mbox{for} \; \Im (k) = 0 \qd.
\eeq
The conjugation matrix $G(k)$ is of dimension two. It is defined only
on the real axis. The matrix $\chi (k)$ should also satisfy the
normalization condition
\beq
     \lim_{k \rightarrow \infty} \chi (k) = I =
        {1 \qd 0 \choose 0 \qd 1} \qd.
\eeq
The problem is to find such a matrix $\chi(k)$. In order to relate
$\chi (k)$ to the integral operator (\ref{kerv}) we shall use a
conjugation matrix $G(k)$ of the special form
\beq \label{gk}
     G(k) = {1 \qd 0 \choose 0 \qd 1} + 2\p \i
            \left( \begin{array}{cc}
	       - e_-(k) e_+(k) & e_+^2(k) \\
	       - e_-^2(k) & e_+(k) e_-(k)
            \end{array} \right) \qd.
\eeq
The Riemann-Hilbert problem with this form of conjugation matrix was
studied in detail in chapter XV of the book \cite{KBIBo}. Let us
recall some fundamental facts:

One can prove that
\beq
     {f_+(k) \choose f_-(k)} = \chi (k) {e_+(k) \choose e_-(k)} \qd.
\eeq
The functions $f_\pm (k)$ were defined by the integral equations
(\ref{fies}). On the other hand, the Rie\-mann-Hilbert problem can be
solved by means of these functions
\beq \label{rhsol}
     \chi (k) = {1 \qd 0 \choose 0 \qd 1} + \int_{- \infty}^\infty \!
                \frac{dp}{p - k} \left( \begin{array}{cc}
		   e_-(p) f_+(p) & - e_+(p) f_+(p) \\
		   e_-(p) f_-(p) & - e_+(p) f_-(p)
                \end{array} \right) \qd.
\eeq
In order to express the potentials $B_{ab}$ and $C_{ab}$ in terms of
$\chi (k)$, let us study the asymptotics of $\chi (k)$ as $k \rightarrow
\infty$,
\beq \label{rhas}
     \chi (k) = I + \frac{\ps_1}{k} + \frac{\psi_2}{k^2}
                            + \CO \left( \frac{1}{k^3} \right) \qd.
\eeq
Comparing (\ref{rhsol}), (\ref{rhas}) and (\ref{defpot}) we obtain
\beq
     \psi_1 =  {- B_{-+} \qd B_{++} \choose - B_{--} \qd B_{+-}}
        \qd, \qd
     \psi_2 = {- C_{-+} \qd C_{++} \choose - C_{--} \qd C_{+-}} \qd.
\eeq
In such a way we have formulated the Riemann-Hilbert problem and related
it to the integral operator $\hat V$ and to the correlation functions
$G_\auf^+ (x,t)$ and $G_\auf^- (x,t)$. Next we shall solve the
Riemann-Hilbert problem at large space time separation $x \rightarrow
\infty$, $t \rightarrow \infty$, $x/t$ fixed. Then we shall calculate
the potentials $B_{ab}$ and $C_{ab}$ from (\ref{rhas}). The knowledge
of the potentials will permit us to evaluate $\det(\hat I + \g \hat V)$,
see (\ref{dxtdet}). This will give us an asymptotic expression for
the correlation functions (\ref{gp2}), (\ref{gm2}). We also want to
note that
\beq
     \det \, G(k) = \det \, \chi (k) = 1 \qd.
\eeq
\section{Asymptotic solution of the Riemann-Hil\-bert problem}
In order to solve the Riemann-Hilbert problem asymptotically
\cite{IIKV92,Manakov74,ZaMa76,Its81,ItNo86,DIZ93,DeZh93}, we shall
reformulate it twice. This will bring it into the canonical form, which
was solved by S.~V.~Manakov \cite{Manakov74,ZaMa76}.

First reformulation: We want to simplify the conjugation matrix $G(k)$.
Let us define a matrix
\beq \label{chizero}
     \chi_0 (k) = \left( \begin{array}{cc}
                     1 & \dst{- (1 - \cos(\h))
		     \int_{- \infty}^\infty \frac{dp \, e^{- \tau(p)}}
		     {2 \p(p - k)}} \\
		     0 & 1 \end{array} \right) \qd.
\eeq
Here $\tau(p) = \i p^2 t - \i p x$. The matrix $\chi_0 (k)$ is analytic 
for $\Im(k) > 0$ and for $\Im(k) < 0$. At large $k$ it approaches the
unit matrix as a Taylor series in $1/k$. Now, instead of the unknown
matrix function $\chi (k)$, let us introduce a new unknown matrix
function $\tilde \chi (k)$,
\beq \label{chitilde}
     \chi (k) = \tilde \chi (k) \chi_0 (k) \qd.
\eeq
Let us formulate the Riemann-Hilbert problem for $\tilde \chi (k)$.
\begin{enumerate}
\item
$\tilde \chi (k)$ is analytic for $\Im(k) > 0$ and for $\Im(k) < 0$.
\item
As $k \rightarrow \infty$
\beq \label{chitias}
     \tilde \chi (k) = I + \frac{\tilde \chi_1}{k}
                         + \frac{\tilde \chi_2}{k^2}
			 + \CO \left( \frac{1}{k^3} \right) \qd.
\eeq
\item
$\tilde \chi (k)$ has a discontinuity across the real axis, which is
described by a conjugation matrix $\tilde G(k)$,
\beq
     \tilde \chi_- (k) = \tilde \chi_+ (k) \tilde G(k) \qd
        \mbox{for} \; \Im (k) = 0 \qd.
\eeq
\end{enumerate}
Let us calculate
\beq \label{gt}
     \tilde G(k) = \left( \begin{array}{cc}
                   \tilde G_{11} (k) & \tilde G_{12} (k) \\
                   \tilde G_{21} (k) & \tilde G_{22} (k)
		   \end{array} \right) =
                   \chi_0^+ (k) G(k) (\chi_0^- (k))^{- 1} \qd.
\eeq
Substituting (\ref{gk}) and (\ref{chizero}) into (\ref{gt}) we obtain
\beq \label{gtilde}
     \begin{array}{l@{\: , \:}l}
        \tilde G_{11} (k) = 1 + \g \dh(k) (e^{- \i \h} - 1) &
        \tilde G_{12} (k) = - \i (1 - \cos(\h))
                            e^{- \tau(k)}(1 - \g \dh(k)) \: , \\
        \tilde G_{21} (k) = -2 \i \g \dh (k) e^{\tau(k)} &
        \tilde G_{22} (k) = 1 + \g \dh(k) (e^{\i \h} - 1) \: ,
     \end{array}
\eeq
where
\beq
     \g \dh(k) = \frac{2 \cosh(B/T)}{2 \cosh(B/T) + e^{(k^2 - \m)/T}}
                 \qd.
\eeq
Now the conjugation matrix is an elementary function of $k$,
$\det \, \tilde G(k) = 1$ and $\det \, \tilde \chi (k) = 1$.

Using (\ref{chizero}) and (\ref{chitilde}) we can also evaluate the
coefficients of the expansion (\ref{chitias}) of $\tilde \chi (k)$
for large $k$,
\beq
     \tilde \chi_1 = {- B_{-+} \qd b_{++} \choose - B_{--} \qd B_{+-}}
\eeq
and
\beq
     \tilde \chi_2 = \left( \begin{array}{cc}
                        - C_{-+} & C_{++} + (1 - \cos(\h))
			(\i \6_x G + G B_{+-}) \\
			- C_{--} & C_{+-} + (1 - \cos(\h)) G B_{--}
                     \end{array} \right) \qd.
\eeq
Let us recall that
\beq
     G(x,t) = \frac{1}{2\p} \int_{- \infty}^\infty \! dk \,
              e^{- \tau(k)} = \frac{e^{- \i \p /4}}{2 \sqrt{\p t}}
	      e^{\i x^2/4t} \qd.
\eeq
In order to prepare for the second step let us introduce a function
$\a(k)$,
\beq
     \a(k) = \exp \left\{ \frac{\i}{2 \p} \int_{- \infty}^\infty
             \frac{dp}{p - k} \ln \left(1 + \g \dh(p)(e^{- \i \h} - 1)
	     \right) \right\} \qd.
\eeq
This function solves the scalar Riemann-Hilbert problem
\beq
     \a_- (k) = \a_+ (k) \left( 1 + \g \dh(k) (e^{- \i \h} - 1) \right)
                \qd.
\eeq
The conjugation function here is the element $\tilde G_{11} (k)$ of
the conjugation matrix $\tilde G (k)$, eq.\ (\ref{gtilde}). Later we
shall need the coefficients $\a_1$, $\a_2$ of the expansion of
$\ln(\a(k))$ for large $k$,
\beq \label{alphasy}
     \ln(\a(k)) = \frac{\a_1}{k} + \frac{\a_2}{k^2}
                  + \CO \left( \frac{1}{k^3} \right) \qd.
\eeq

Second reformulation: We want to bring the Riemann-Hilbert problem to
the canonical form. Instead of $\tilde \chi (k)$ let us define another
$2 \times 2$-matrix
\beq
     \PH (k) = \tilde \chi (k) e^{- \s^z \ln \a(k)} \qd.
\eeq
This matrix function is analytic for $\Im(k) > 0$ and $\Im(k) < 0$. It
approaches the unit matrix at large $k$,
\beq \label{phasy}
     \PH (k) = I + \frac{\PH_1}{k} + \frac{\PH_2}{k^2} + 
               \CO \left(\frac{1}{k^3} \right) \qd.
\eeq
It has a discontinuity on the real axis, which can be described by
a conjugation matrix $G_\PH (k)$,
\beq \label{phirh}
     \PH_- (k) = \PH_+ (k) G_\PH (k) \qd.
\eeq
Here
\beq
     G_\PH (k) = e^{\s^z \ln \a_+ (k)} \tilde G(k)
                 e^{- \s^z \ln \a_-(k)} \qd.
\eeq
The matrix elements of $G_\PH (k)$ are
\beq
     \begin{array}{l@{\qd, \qd}l}
        (G_\PH)_{11} = 1 &
	(G_\PH)_{12} = \a_+(k) \a_-(k) \tilde G_{12} (k) \qd, \\
	\dst{(G_\PH)_{21} = \frac{\tilde G_{21} (k)}{\a_-(k) \a_+(k)}} &
	(G_\PH)_{22} = \tilde G_{11} (k) \tilde G_{22} (k) \qd.
     \end{array}
\eeq
We also note that
\beq
     \det \, G_\PH (k) = \det \, \PH (k) = 1 \qd.
\eeq
Let us introduce the notations
\bea \label{pandq}
     p(k) & = & - \i (1 - \cos(\h))(1 - \g \dh(k)) \a_+(k) \a_-(k) \qd,
                \\
     q(k) & = & \frac{- 2 \i \g \dh(k)}{\a_-(k) \a_+(k)} \qd.
\eea
Then we can write the conjugation matrix $G_\PH (k)$ in the canonical
form
\beq \label{gphi}
     G_\PH (k) = \left( \begin{array}{cc}
                 1 & p(k) e^{- \tau(k)} \\
		 q(k) e^{\tau(k)} & 1 + p(k) q(k)
		 \end{array} \right) \qd.
\eeq
Here $\tau(k) = \i k^2 t - \i kx$. In the following the stationary
point
\beq
     k_0 = \frac{x}{2t}
\eeq
of the phase $\tau(k)$, $\6_k \tau (k_0) = 0$, will play an important
role.

Let us express the potentials $B_{ab}$ in terms of $\PH_1$, eq.\
(\ref{phasy}),
\beq \label{ph1}
     \PH_1 = \left( \begin{array}{cc}
                - B_{+-} - \a_1 & b_{++} \\
                - B_{--} & B_{+-} + \a_1
	     \end{array} \right) \qd.
\eeq
Here
\beq \label{alpha1}
     \a_1 = \frac{1}{2 \p \i} \int_{- \infty}^\infty \! dp \,
            \ln \left( 1 + \g \dh(p) (e^{- \i \h} - 1) \right)
\eeq
is the first coefficient of the expansion (\ref{alphasy}) of 
$\ln(\a(k))$ for large $k$. Using (\ref{dxtdet}) and (\ref{ph1}) we
obtain
\beq
     \6_x \, \ln \det \left( \hat I + \g \hat V \right)
        = \i B_{+-} = \i B_{-+} = - \i ((\PH_1)_{11} + \a_1) \qd.
\eeq
The time derivative of $\ln \det \left( \hat I + \g \hat V \right)$
can be expressed in terms of $\PH_2$, eq. (\ref{phasy}),
\bea \nn
     \6_t \, \ln \det \left( \hat I + \g \hat V \right)
        & = & - \i (C_{+-} +  C_{-+} + (1 - \cos(\h))G B_{--}) \\
        & = & \i \left\{ \tr(\s^z \PH_2) + 2 \a_2 \right\} \qd,
\eea
where $\a_2$ is the second coefficient in the expansion (\ref{alphasy}),
\beq \label{alpha2}
     \a_2 = \frac{1}{2 \p \i} \int_{- \infty}^\infty \! dp \,
            p \ln \left( 1 + \g \dh(p) (e^{- \i \h} - 1) \right) \qd.
\eeq
This is the end of the second reformulation.

The asymptotic solution of the Riemann-Hilbert problem with conjugation
matrix $G_\PH (k)$, eq.\ (\ref{gphi}), is known. It is given by
Manakov's Ansatz (see page 452 of the book \cite{KBIBo}). In order
to describe Manakov's Ansatz we first solve a scalar Riemann-Hilbert
problem for a function $\de(k)$,
\beq
     \de_+(k) = \de_-(k) \left( 1 + p(k) q(k) \th(k_0 - k) \right) \qd.
\eeq
Here $\th$ is the Heavyside step function,
\beq
     \th (k) \, = \, \left\{ {1 \atop 0} \right. \qd \mbox{if} \qd
                     {k > 0 \atop k < 0} \qd.
\eeq
The solution is
\beq \label{delta}
     \de (k) = \exp \left\{ \frac{1}{2 \p \i} \int_{- \infty}^{k_0}
               \frac{du}{u - k} \ln(1 + p(u) q(u)) \right\} \qd.
\eeq
Now we can define functions $I^p(k)$ and $I^q(k)$,
\bea
     I^p(k) & = & \frac{1}{2 \p \i} \int_{- \infty}^\infty
                  \frac{du}{u - k} \; \de_+(u) \de_-(u)
		  p(u) e^{- \tau(u)} \qd, \\
     I^q(k) & = & \frac{1}{2 \p \i} \int_{- \infty}^\infty
                  \frac{du}{u - k} \; \frac{q(u) e^{\tau(u)}}
		  {\de_+(u) \de_-(u)} \qd.
\eea
Using these functions we can write the solution of the Riemann-Hilbert
problem (\ref{phirh}), (\ref{gphi}) as
\beq \label{phmanakov}
     \PH (k) = \left( \begin{array}{cc}
               1 & - I^p (k) \\ - I^q (k) & 1
	       \end{array} \right) e^{\s^z \ln \de(k)}
	       + \CO \left( t^{- \2} \right) \qd.
\eeq
This solution is an asymptotic solution which is valid only as
$x \rightarrow \infty$, $t \rightarrow \infty$ for fixed ratio $x/t$.
\section{Asymptotics of the correlation functions}
We solved the Riemann-Hilbert problem. Now we can calculate the
logarithmic derivatives of the determinant,
\bea \label{dxdet3}
        \6_x \, \ln \det \left( \hat I + \g \hat V \right)
	   & = & \i B_{+-} = \i B_{-+} = - \i ((\PH_1)_{11} + \a_1)
	   \qd, \\
     \label{dtdet3}
        \6_t \, \ln \det \left( \hat I + \g \hat V \right)
	   & = & \i \left\{ \tr( \s^z \PH_2 ) + 2 \a_2 \right\} \qd.
\eea
In order to obtain $\PH_1$ and $\PH_2$ we have to decompose our
asymptotic solution (\ref{phmanakov}) into a Taylor series in $1/k$
(\ref{phasy}). Then
\beq
     -B_{+-} = \a_1 + \de_1 \qd.
\eeq
For $\a_1$ see (\ref{alpha1}). In order to define $\de_1$ let us
recall that
\beq
     \ln(\de (k)) = \frac{1}{2 \p \i} \int_{- \infty}^{k_0}
               \frac{dp}{p - k} \ln \left(1 - 2 (1 - \cos(\h)) \g \dh(p)
	       (1 - \g \dh(p))\right) \qd,
\eeq
$k_0 = x/2t$. Here we used (\ref{pandq}) and (\ref{delta}). At large
$k$, $\ln(\de(k))$ can be decomposed into a Taylor series in $1/k$,
\beq
     \ln(\de(k)) = \frac{\de_1}{k} + \frac{\de_2}{k^2} +
                   \CO \left( \frac{1}{k^3} \right) \qd.
\eeq
So
\bea
     \de_1 & = & - \frac{1}{2 \p \i} \int_{- \infty}^{k_0}
               dp \ln \left(1 - 2 (1 - \cos(\h)) \g \dh(p)
	       (1 - \g \dh(p)) \right) \qd, \\ \label{delta2}
     \de_2 & = & - \frac{1}{2 \p \i} \int_{- \infty}^{k_0}
               dp \, p \ln \left(1 - 2 (1 - \cos(\h)) \g \dh(p)
	       (1 - \g \dh(p)) \right) \qd.
\eea
Substitution of $\a_1$, eq.\ (\ref{alpha1}), and $\de_1$ into
(\ref{dxdet3}) gives
\beq \label{bpmasy}
     B_{+-} = - \frac{1}{2\p \i} \int_{-\infty}^\infty dp \,
              \sign(p - k_0) \ln \left( 1 + \g \dh(p) \left(
	      e^{- \i \h \ssign(p - k_0)} - 1 \right) \right) \qd,
\eeq
where $k_0 = x/2t$ and $\sign(k)$ is the sign function,
\beq
     \sign (k) \, = \, \left\{ \begin{array}{r} 1 \\ -1 \end{array} \right.
		       \qd \mbox{if} \qd
		       \begin{array}{c} k > 0 \\ k < 0 \end{array} \qd.
\eeq
Using(\ref{dxdet3}) we obtain
\bea \nn
     \lefteqn{\6_x \ln \det(\hat I + \g \hat V) =} \\ \label{dxdet4}
              && - \frac{1}{2\p}
              \int_{-\infty}^\infty dp \,
              \sign(p - k_0) \ln \left( 1 + \g \dh(p) \left(
	      e^{- \i \h \ssign(p - k_0)} - 1 \right) \right) \qd.
\eea
We shall calculate the time derivative in a similar way,
\beq
     \tr(\s^z \PH_2) = 2 \de_2 \qd,
\eeq
see (\ref{phmanakov}) and (\ref{delta2}). From (\ref{dtdet3}) we have
\beq
     \6_t \ln \det(\hat I + \g \hat V) = 2 \i (\de_2 + \a_2) \qd,
\eeq
see (\ref{alpha2}) and (\ref{delta2}). Finally,
\bea \nn
     \lefteqn{\6_t \ln \det(\hat I + \g \hat V) =} \\ && \frac{1}{\p}
              \int_{-\infty}^\infty dp \, p \,
              \sign(p - k_0) \ln \left( 1 + \g \dh(p) \left(
	      e^{- \i \h \ssign(p - k_0)} - 1 \right) \right) \qd.
\eea
Now we have calculated space (\ref{dxdet4}) and time derivative of
$\ln \det(\hat I + \g \hat V)$. Let us integrate,
\beq \label{detasy1}
    \ln \det(\hat I + \g \hat V) = \frac{1}{2 \p}
              \int_{-\infty}^\infty dp \, |x - 2pt|
	      \ln \left( 1 + \g \dh(p) \left(
	      e^{- \i \h \ssign(p - k_0)} - 1 \right) \right) \qd.
\eeq
This is the leading term of the asymptotics. Let us recall that
\beq
     \g \dh(k) = \frac{2 \cosh(B/T)}{2 \cosh(B/T) + e^{(k^2 - \m)/T}}
                 \qd.
\eeq

In order to evaluate the correlation functions (\ref{gp2}), (\ref{gm2})
we also need $b_{++}$ and $B_{--}$.
\beq
     b_{++} = (\PH_1)_{12} \qd, \qd B_{--} = - (\PH_1)_{21} \qd,
\eeq
see (\ref{ph1}). Using the explicit estimate for the error (see page
453 of the book \cite{KBIBo}) of the asymptotic solution $\PH(k)$,
eq.\ (\ref{phmanakov}), of the Riemann-Hilbert problem we find
\beq
     |b_{++}| \sim \frac{1}{\sqrt{t}} \qd, \qd
     |B_{--}| \sim \frac{1}{\sqrt{t}} \qd,
\eeq
which shows that $b_{++}$ and $B_{--}$ decay with time. In order to
get a more precise estimate of the asymptotics, we shall turn to the
differential equations (\ref{deqs}). We shall use some information
from the book \cite{KBIBo}. So we shall rewrite the differential
equations (\ref{deqs}) in the variables $\tilde x = - x/2$,
$\tilde t = t/2$. Then $k_0 = - \tilde x /2 \tilde t$, and
\beq \label{tdeqs}
     \begin{array}{rcl}
      - \;  \i \, \dst{\frac{\6 b_{++}}{\6 \tilde t}} & = &
           \dst{\2 \; \frac{\6^2 b_{++}}{\6 \tilde x^2}}
           + 4 b_{++}^2 B_{--} \qd, \\[2ex]
        \i \, \dst{\frac{\6 B_{--}}{\6 \tilde t}} & = &
	   \dst{\2 \; \frac{\6^2 B_{--}}{\6 \tilde x^2}}
           + 4 b_{++} B_{--}^2 \qd.
     \end{array}
\eeq
The asymptotics of the decaying solutions of this system is well known
\cite{SeAb76,AbSeBo,BuSu82}. The leading terms are
\beq \label{bppmmasy}
     \begin{array}{r@{\: = \:}l}
        b_{++} & \dst{\tilde u_0 \, {\tilde t}^{- 1/2 - \i \nu}
	         e^{\i \tilde x^2/2 \tilde t}} \qd, \\[2ex]
        B_{--} & \dst{\tilde v_0 \, {\tilde t}^{- 1/2 + \i \nu}
	         e^{- \i \tilde x^2/2 \tilde t}} \qd.
     \end{array}
\eeq
Here $\tilde u_0$ and $\tilde v_0$ only depend on $k_0$, $T$,
$\m$ and $B$. Substitution of these asymptotic solutions into the
differential equations (\ref{tdeqs}) gives
\beq \label{nuuv}
     \nu = - 4 \tilde u_0 \tilde v_0 \qd.
\eeq
In order to calculate $\nu$ we use the identity (\ref{dxbpm}),
\beq \label{tdxbpm}
     \frac{\6 B_{+-}}{\6 \tilde x} = 2 \i B_{--} b_{++} \qd.
\eeq
Let us substitute (\ref{bpmasy}) here,
\beq \label{tdxbpm2}
     \frac{\6 B_{+-}}{\6 \tilde x} = \frac{\i}{4 \p \tilde t}
        \ln\left\{ 1 - 2(1 - \cos(\h)) \g \dh(k_0)(1 - \g \dh(k_0))
	   \right\} \qd.
\eeq
{From} (\ref{bppmmasy}) and (\ref{nuuv}) we have
\beq
     2 \i B_{--} b_{++} = \frac{2 \i \tilde u_0 \tilde v_0}{\tilde t}
        = \frac{- \i \nu}{2 \tilde t} \qd.
\eeq
Using (\ref{tdxbpm}), (\ref{tdxbpm2}) we get
\beq
     \nu = - \frac{1}{2 \p} \ln\left\{
             1 - 2(1 - \cos(\h)) \g \dh(k_0)(1 - \g \dh(k_0))
	     \right\} \qd.
\eeq
So we have calculated the asymptotics of $b_{++}$ and $B_{--}$, see
(\ref{bppmmasy}). We can use the differential equations (\ref{tdeqs})
even more. We can improve the asymptotic expression (\ref{detasy1})
for $\ln \det(\hat I + \g \hat V)$. Following the steps from pages
455-457 of the book \cite{KBIBo} we obtain
\bea \label{detasy2}
    \lefteqn{\ln \det(\hat I + \g \hat V) =} \nn \\ &&
              \frac{1}{2 \p}
              \int_{-\infty}^\infty dp \, |x - 2pt|
	      \ln \left( 1 + \g \dh(p) \left(
	      e^{- \i \h \ssign(p - k_0)} - 1 \right) \right)
	      + \frac{\nu^2}{2} \ln t \: . \qd
\eea
I order to get the correlation functions (\ref{gp2}), (\ref{gm2}) let
us calculate
\bea
     \lefteqn{b_{++} \det(\hat I + \g \hat V) =
        u_0 t^{- \2 (1 + \i \nu)^2} \exp(\i x^2/4t)} \nn \\ && \cdot
              \exp \left\{ \frac{1}{2 \p}
              \int_{-\infty}^\infty dp \, |x - 2pt|
	      \ln \left( 1 + \g \dh(p) \left(
	      e^{- \i \h \ssign(p - k_0)} - 1 \right) \right) \right\}
	      \: , \qd \\
     \lefteqn{B_{--} \det(\hat I + \g \hat V) =
        v_0 t^{- \2 (1 - \i \nu)^2} \exp(- \i x^2/4t)} \nn \\ && \cdot
              \exp \left\{ \frac{1}{2 \p}
              \int_{-\infty}^\infty dp \, |x - 2pt|
	      \ln \left( 1 + \g \dh(p) \left(
	      e^{- \i \h \ssign(p - k_0)} - 1 \right) \right) \right\}
	      \: . \qd
\eea
We did not calculate the coefficients $u_0$ and $v_0$ in these
expressions. We calculated the leading exponential factor and power
(of time) corrections. Further down, while substituting we shall not
pay attention to numerical coefficients. Let us get all notations
together, and let us present our asymptotic expressions for the
correlation functions (\ref{gp1}), (\ref{gm1}),
\bea \label{gp4}
     G_\auf^+ (x,t) & = & e^{\i t(\m - B) + i x^2/4t}
        \oint \frac{dz}{(z - 1)^2} \, u_0 F(z)
	      t^{- \2 (1 + \i \nu (z))^2}
	      e^{t S (z)} \qd, \\ \label{gm4}
     G_\auf^- (x,t) & = & e^{- \i t(\m - B) - i x^2/4t}
        \oint \frac{dz}{z} \, v_0 F(z) t^{- \2 (1 - \i \nu (z))^2}
	      e^{t S (z)} \qd.
\eea
Here the contour of integration is the unit circle.
\bea
     F(z) & = & 1 + \frac{z}{\g - z} + \frac{1}{\g z - 1} \qd, \\
     \nu (z) & = & - \frac{1}{2 \p} \ln\left\{
             1 - (2 - z - z^{-1}) \g \dh(k_0)(1 - \g \dh(k_0))
	     \right\} \qd, \\
     t S (z) & = &
	      \frac{1}{2 \p}
              \int_{-\infty}^\infty dp \, |x - 2pt|
	      \ln \left( 1 + \g \dh(p) \left(
	      z^{- \ssign(p - k_0)} - 1 \right) \right) \qd, \\
     \g & = & 1 + e^{2B/T} \: , \: k_0 = x/2t \: , \:
     \g \dh(k) = \frac{2 \cosh(B/T)}{2 \cosh(B/T) + e^{(k^2 - \m)/T}}
                 \: . \qd
\eea
At the end of section 2 we showed that $b_{++}$ cancels the second
order pole at $z = 1$. This means that $u_0$ will cancel the second
order pole at $z = 1$ in (\ref{gp4}). The remaining contour integrals in
(\ref{gp4}), (\ref{gm4}) can be evaluated by the method of steepest
descent (see next section). Eqs.\ (\ref{gp4}), (\ref{gm4}) are
integral formulae for the large time, long distance asymptotics of
correlation functions of local fields for the delta-interacting
electron gas. These formulae are valid for arbitrary finite temperatures.
They are the main result of the present article. In the next section
we shall simplify our asymptotic expressions (\ref{gp4}), (\ref{gm4})
at low temperatures. Let us note that $S(z)$ can be written as
\beq
     S (z) = \frac{1}{\p} \int_{-\infty}^\infty dp \,
               |p - k_0| \ln \left\{
	       \frac{e^{(k^2 + B - \m)/T} + \g z^{- \ssign(p - k_0)}}
	       {e^{(k^2 + B - \m)/T} + \g} \right\} \qd.
\eeq
\section{Low temperature asymptotics}
We obtained the asymptotic formulae (\ref{gp4}), (\ref{gm4}). The
remaining contour integral should be taken by means of the steepest
descent (saddle point) method. Let us consider here only the leading
asymptotic factor. Then we can rewrite the simplified expression
for the asymptotics as
\bea
     G_\auf^+ (x,t) & = & e^{\i t (\m - B)} e^{\i x^2/4t} \, \Om_+
                          \qd, \\
     G_\auf^- (x,t) & = & e^{- \i t (\m - B)} e^{- \i x^2/4t} \, \Om_-
                          \qd, \\ \label{cald}
     \Om_\pm & = & \oint dz \, F(z) t^{- \2 (1 \pm \i \nu (z))^2}
                   e^{t S(z)} \qd.
\eea
The saddle point equation $\6 S/\6 z = 0$ can be represented in the
form
\beq \label{saddle}
     \int_0^\infty \frac{dk \, k}{1 + \frac{1}{\g z}
        e^{((k - k_0)^2 + B - \m)/T}} =
     \int_0^\infty \frac{dk \, k}{1 + \frac{z}{\g}
        e^{((k + k_0)^2 + B - \m)/T}} \qd,
\eeq
$k_0 = x/2t$, $\g = 1 + e^{2B/T}$. We prove in appendix A that there
is only one solution of this equation in the interval $0 < z \le 1$.
Considering the pure time direction, $k_0 = 0$, we find the solution
\beq
     z = 1 \qd,
\eeq
which leads to
\beq
     \Om_\pm \sim t^{-1} \qd.
\eeq
This result is valid for all finite temperatures.

Let us consider the other asymptotic regions, $k_0 > 0$,
$x \rightarrow \infty$ and $t \rightarrow \infty$. In these regions
the saddle point equation (\ref{saddle}) can be solved explicitly
only for small temperatures. There are two solutions $z_\pm = \pm z_c$,
\beq
     z_c = \frac{T^{3/4}}{2 \p^{1/4} k_0^{3/2}} \, e^{- k_0^2/2T} \qd.
\eeq
If we deform the contour of integration from a circle to a line
through $z_+$ parallel to the imaginary axis, we pass the saddle
point in the right direction (see appendix A).

In appendix A we obtain the following low temperature expansion for
the phase $tS(z)$,
\beq \label{lowphase}
     t S(z) = - 2 k_0 D t \left\{ \left( 1 - \frac{1}{z} \right)
	          z_c^2 + 1 - z \right\} \qd.
\eeq
Here $D$ is the density (\ref{lowtdens}),
\beq
     D \; = \; \left\{ \begin{array}{l}
               \sqrt{\frac{T}{\p}} \, e^{\m/T} \\[1ex]
               \frac{1}{2} \sqrt{\frac{T}{\p}} \, e^{(\m + |B|)/T}
	       \end{array} \right.
	       \qd \mbox{if} \qd
	       \begin{array}{l} B = 0 \qd, \\[1ex] B \ne 0 \qd.
	       \end{array}
\eeq
Eq.\ (\ref{lowphase}) shows that at low temperatures the relevant
parameter for the calculation of the asymptotics of $\Om_\pm$ is
$2k_0Dt = xD$ rather than $t$. The parameter $xD$ has a simple
interpretation. It is the average number of particles in the interval
$[0,x]$. We have to distinguish two limiting cases.
\begin{enumerate}
\item
$xD \rightarrow 0$, the number of electrons in the interval $[0,x]$
vanishes. In this regime the interaction of the electrons is
negligible. An electron propagates freely from 0 to $x$. $\Om_\pm$
cannot be calculated by the method of steepest descent. We have to
use (\ref{gp4}), (\ref{gm4}) to calculate the asymptotics of
$G_\auf^+ (x,t)$ and $G_\auf^- (x,t)$. Now $tS(z)$ and $\nu(z)$ tend
to zero on the contour of integration. Thus $G_\auf^\pm (x,t) \sim
t^{- \2} e^{\pm \i t(\m - B) \pm \i x^2/4t}$, which, as expected, is
the same as for free fermions. The proportionality factor is different
for $G_\auf^+ (x,t)$ and $G_\auf^- (x,t)$. To calculate it from
(\ref{gp4}), (\ref{gm4}) we would have to know $u_0$ and $v_0$. By
comparison with free fermions (see below) we expect it to be a
number for $G_\auf^+ (x,t)$ and proportional to $e^{(\m - B -k_0^2)/T}$
for $G_\auf^- (x,t)$.
\item
$xD \gg 1$, the average number of electrons in the interval $[0,x]$
is large. Now the interaction becomes important, and we can use the
method of steepest descent to calculate $\Om_\pm$. This is the most
interesting case. In the following we will investigate it in detail.
\end{enumerate}

The contribution of the saddle point to $\Om_\pm$, eq.\ (\ref{cald}),
is
\beq \label{edens}
     \Om_\pm \sim t^{- (1 \pm \i \nu(z_+))} \, e^{- xD_\ab} \qd,
\eeq
where
\bea \label{nuzp}
     \nu(z_+) & = & - \; \frac{2D_\ab k_0^{3/2} e^{- k_0^2/2T}}
                          {\p^{1/4} T^{5/4}} \qd, \\
     D_\ab & = & \tst{\2 \sqrt{\frac{T}{\p}} e^{(\m + B)/T}} \qd.
\eea
$D_\ab$ is the low temperature expression for the density of down-spin
electrons, eq.\ (\ref{densab}).

We further have to take into account the pole of the function $F$
at $z = \g^{-1}$. While deforming the contour, we could have crossed it.
It turns out that the pole contributes to $\Om_\pm$, when the magnetic
field $B$ is below a critical positive value
\beq \label{bc}
     B < B_c = k_0^2/4
\eeq
(see appendix A). The pole contribution always dominates the
contribution of the saddle point. It is obtained as
\bea \label{omum}
     \Om_\pm & \sim & t^{- \left(\2 \pm \i \nu(\g^{-1}) \right)}
                  e^{- x D_\ab} \qd, \\ \label{nugamma}
     \nu(\g^{-1}) & = & - \; \frac{e^{(3B + \m - k_0^2)/T}}{2\p} \qd.
\eea
Finally, we have the following low temperature asymptotic expressions
for the leading factors of the correlation functions. For
\bea
     B < B_c : \nn \\ \label{gpas}
     G_\auf^+ (x,t) & = & t^{- \2 - \i \nu(\g^{-1})}
                          e^{\i t (\m - B)} e^{\i x^2/4t} e^{- x D_\ab}
			  \qd, \\ \label{gmas}
     G_\auf^- (x,t) & = & t^{- \2 + \i \nu(\g^{-1})}
                          e^{- \i t (\m - B)} e^{- \i x^2/4t}
			  e^{- x D_\ab} \qd,
\eea
where $\nu(\g^{-1})$ is given by eq.\ (\ref{nugamma}), and for
\bea
     B > B_c : \nn \\ \label{gpas2}
     G_\auf^+ (x,t) & = & t^{- 1 - \i \nu(z_+)}
                          e^{\i t (\m - B)} e^{\i x^2/4t} e^{- x D_\ab}
			  \qd, \\ \label{gmas2}
     G_\auf^- (x,t) & = & t^{- 1 + \i \nu(z_+)}
                          e^{- \i t (\m - B)} e^{- \i x^2/4t}
			  e^{- x D_\ab} \qd,
\eea
where $\nu(z_+)$ is given by eq.\ (\ref{nuzp}). Recall that we have
assumed the average number of electrons in the interval $[0,x]$ to be
large, $xD \gg 1$. The corresponding expressions for $G_\ab^\pm (x,t)$
follow from equation (\ref{aufab}).

The leading low temperature expressions for $D$ and $D_\ab$ agree for
positive magnetic field. Hence, above the critical field $B_c$ the
correlation functions depend on the magnetic field only through the
trivial factors $e^{\pm \i t(\m - B)}$. The system is saturated. Another
way of looking at the condition (\ref{bc}), which would be appropriate
for experiments with fixed magnetic field $B$, is the following. For
positive magnetic field there are two different asymptotic regions.
One is the space like region, where $k_0^2 > 4B$, the other one is the
time like region, where $k_0^2 < 4B$. Correlation functions of local
fields have asymptotics (\ref{gpas}), (\ref{gmas}) in the space like
region and asymptotics (\ref{gpas2}), (\ref{gmas2}) in the time like
region. For non-positive magnetic field there is no distinction
between space like and time like region.

It is instructive to compare the asymptotic expressions (\ref{gpas}) -
(\ref{gmas2}) for the correlation functions with the corresponding
asymptotics for free fermions. The two-point correlation functions for
free fermions have the low temperature asymptotics
\bea \label{gpf}
     \frac{\tr \left( e^{- H_0/T} \,
        \ps_\auf (x,t) \ps_\auf^\dagger (0,0) \right)}
	{\tr \left( e^{- H_0/T} \right)}
     & = & C_+ \, t^{- \2} e^{\i t (\m - B)} e^{\i x^2/4t} \qd, \\
        \label{gmf}
     \frac{\tr \left( e^{- H_0/T} \,
        \ps_\auf^\dagger (x,t) \ps_\auf (0,0) \right)}
	{\tr \left( e^{- H_0/T} \right)}
     & = & C_- \, t^{- \2} e^{- \i t (\m - B)} e^{- \i x^2/4t} \qd,
\eea
as $x \rightarrow \infty$, $t \rightarrow \infty$. Here $H_0$ is the
free Hamiltonian ((\ref{ham}) with $c = 0$). The leading low temperature
contributions to the constants in (\ref{gpf}) and (\ref{gmf}) are
$C_+ = e^{- \i \frac{\p}{4}}/2 \sqrt{\p}$ and $C_- =
e^{(\m - B -k_0^2)/T} e^{\i \frac{\p}{4}}/2 \sqrt{\p}$. Comparing the
free fermionic expressions with (\ref{gpas}), (\ref{gmas}) or
(\ref{gpas2}), (\ref{gmas2}), respectively, we can interpret the factors
$t^{\pm \i \nu(\g^{-1})} e^{- x D_\ab}$ and $t^{- \2 \pm \i \nu(z_+)}
e^{- x D_\ab}$ as low temperature corrections, which appear due to the
interaction. The occurrence of the density of down-spin electrons in
the exponential factors in eqs.\ (\ref{gpas}) - (\ref{gmas2}) has
a natural interpretation. Correlations decay due to interaction. Because
of the Pauli principle and the locality of the interaction in our
specific model, eq.\ (\ref{ham}), up-spin electrons interact only with
down-spin electrons. Therefore the correlation length is expected to
be a decreasing function of the density of down-spin electrons,
which diverges as the density of down spin electrons goes to zero.
We see from (\ref{gpas}) - (\ref{gmas2}) that the low temperature
expression for the correlation length is just $1/D_\ab$ and thus
meets these expectations.

In the limit $B \rightarrow - \infty$, $\m \rightarrow - \infty$,
$\m - B$ fixed there are no $\ab$-spin electrons left in the system,
$D_\ab \rightarrow 0$, $D_\auf \rightarrow D$.  This is the free
fermion limit. In the free fermion limit $B < B_c$, and the asymptotics
of $G_\auf^+ (x,t)$ and $G_\auf^- (x,t)$ are given by eqs.\
(\ref{gpas}), (\ref{gmas}), which then turn into the expressions
(\ref{gpf}), (\ref{gmf}) for free fermions.

So we have explicitly evaluated the asymptotics of correlation
functions of local fields $G_\auf^\pm (x,t)$, (\ref{gp1}),
(\ref{gm1}). Our main results are formulae (\ref{gp4}), (\ref{gm4}).
In this section we considered the saddle point calculation of the
remaining contour integral, and we discussed the physical meaning of
our asymptotics in the low temperature regime.

\section*{Acknowledgements}
We would like to thank Alexander R.\ Its for helpful discussions.
This work was partially supported by the National Science Foundation
under grant number PHY-9605226 (V.K.), by the Deutsche
Forschungsgemeinschaft under grant number Go 825/2-1 (F.G.)
and by the grant INTAS-RFBR 95-0414 (A.I. and A.P.).

\setcounter{section}{0} 

\renewcommand{\thesection}{\Alph{section}}

\renewcommand{\theequation}{\thesection.\arabic{equation}}

\append{Steepest descent calculation of the leading asymptotic factors}
In this appendix we discuss how to calculate the leading asymptotic
contribution to the integrals in (\ref{gp4}), (\ref{gm4}) by means of
the method of steepest descent. The leading contributions determine
the exponential decay and the power law behavior of the correlation
functions (\ref{gp1}), (\ref{gm1}). We will neglect all sub-leading
factors. Then we are left with the calculation of the contour
integral (\ref{cald}),
\beq \label{leadint}
     \Om_\pm = \oint dz \, F(z) t^{- \2 (1 \pm \i \nu (z))^2}
                           e^{t S(z)} \qd.
\eeq
The contour of integration is the unit circle.
\bea \label{capitalf}
     F(z) & = & 1 + \frac{z}{\g - z} + \frac{1}{\g z - 1} \qd, \\
     \nu (z) & = & - \frac{1}{2 \p} \ln\left\{
             1 - (2 - z - z^{-1}) \g \dh(k_0)(1 - \g \dh(k_0))
	     \right\} \qd, \\
     \label{phaseap}
     S (z) & = & \frac{1}{\p} \int_{-\infty}^\infty dp \,
               |p - k_0| \ln \left\{
	       \frac{e^{(k^2 + B - \m)/T} + \g z^{- \ssign(p - k_0)}}
	       {e^{(k^2 + B - \m)/T} + \g} \right\} \qd.
\eea
Let us cut the complex plane from $- \infty$ to $ - \g^{- 1}
e^{(B - \m)/T}$ and from $- \g e^{(\m - B)/T}$ to zero. The integrand
in eq.\ (\ref{leadint}) is analytic in the cut plane except for the
two simple poles of $F(z)$ at $z = \g^{\pm 1}$. We may therefore deform 
the contour of integration as long as we never cross the cuts and take
into account the contributions from the residua, if we cross $z = \g$
or $z = \g^{-1}$.

Let us comment on the somewhat unusual fact that $t$ in eq.\
(\ref{leadint}) appears not only in the exponent. Calculating an
integral of the form (\ref{leadint}) by the method of steepest descent
amounts to deforming the contour of integration in such a way that
we are left with a Laplace type integral. Let us therefore consider the
large $t$ asymptotics of the integral
\beq \label{intint}
     \int_{- \infty}^\infty dx \, t^{g(x)} e^{- t f(x)} =
     \int_{- \infty}^\infty dx \, e^{- t(f(x) - \e g(x))}
     \qd.
\eeq
Here $\e = \ln(t)/t$ becomes small as $t \rightarrow \infty$. We assume
that $f(x)$ has a unique minimum $x_0$. Then, first calculating the
large $t$ asymptotics of the right hand side of (\ref{intint}) by
Laplace's method and expanding for small $\e$ afterwards, we obtain
the leading asymptotics
\beq
     \int_{- \infty}^\infty dx \, t^{g(x)} e^{- t f(x)} =
     \tst{\sqrt{\frac{2\p}{f''(x_0)}}} \;
     t^{g(x_0) - \2} e^{- t f(x_0)} \qd.
\eeq
This means that the factor $t^{- \2 (1 \pm \i \nu(z))^2}$ in equation
(\ref{leadint}) can be treated the same way as a $t$-independent
function. Thus, to leading order, a saddle point $z_c$ contributes
to $\Om_\pm$ as
\beq
     \Om_\pm \sim t^{- \2 - \2 (1 \pm \i \nu(z_c))^2}
                  e^{- t S(z_c)} \qd.
\eeq

The saddle point equation $\6 S/\6 z = 0$ can be represented in the
form
\beq \label{saddleap}
     \int_0^\infty \frac{dk \, k}{1 + \frac{1}{\g z}
        e^{((k - k_0)^2 + B - \m)/T}} =
     \int_0^\infty \frac{dk \, k}{1 + \frac{z}{\g}
        e^{((k + k_0)^2 + B - \m)/T}} \qd,
\eeq
$k_0 = x/2t$, $\g = 1 + e^{2B/T}$. This is still a transcendental
equation, which cannot be solved explicitly for $z$ except at low or
high temperatures.

Before turning to the case of low temperatures, let us show that
(\ref{saddleap}) restricted to the positive real axis, $z > 0$, has
a unique solution, which is located in the interval $0 < z \le 1$.
Let us define two functions
\beq
     J_\pm (x) = \int_0^\infty \frac{dk \, k}{1 + x e^{(k \pm k_0)^2/T}}
                 \qd.
\eeq
We will further use the abbreviation
\beq
     \La = \frac{e^{(B - \m)/T}}{\g} \qd.
\eeq
Note that $\La > 0$. Using the above conventions the saddle point
equation (\ref{saddleap}) reads
\beq \label{saddleap2}
     f(z) = \frac{J_+ (\La z)}{J_- (\La /z)} = 1 \qd.
\eeq
Now
\beq
     f'(z) = \La \left. \left\{ J_+' (\La z) J_- (\La /z) + (1/z^2)
             J_+ (\La z) J_-' (\La /z) \right\} \right/ J_-^2 (\La /z)
	     \qd,
\eeq
where the primes denote derivatives with respect to the arguments.
Since
\beq
     J_\pm' (x) = - \int_0^\infty dk \, \frac{k e^{(k \pm k_0)^2/T}}
                    {(1 + x e^{(k \pm k_0)^2/T})^2} \, < \, 0
\eeq
and $J_\pm (x) > 0$ for all positive $x$, we conclude that $f(z)$ is
monotonically decreasing for positive $z$.

We can decompose $J_- (x)$ into
\beq
     J_- (x) = 2 k_0 \sqrt{T} I(x) + J_+ (x) \qd,
\eeq
where
\beq
     I(x) = \int_0^\infty dk \, \frac{1}{1 + x e^{k^2}} \qd.
\eeq
It follows that
\beq \label{fofone}
     f(1) = \frac{J_+(\La)}{2 k_0 \sqrt{T} I(\La) + J_+(\La)} \,
            \le \, 1 \qd,
\eeq
since $I(x), J_+(x) > 0$ for all positive $x$. Thus the saddle point
equation (\ref{saddleap}) has no positive real solution $z$ with
$z > 1$. We also learn from (\ref{fofone}) that $z = 1$ solves the
saddle point equation for $k_0 = 0$.

Let us work out the behavior of $f(z)$ for $z \rightarrow 0+$ and
$B$, $\m$, $T$ and $k_0$ fixed. We have the following asymptotic
expansions of $I(x)$ and $J_+(x)$ for large positive $x$,
\bea
     I(x) & = & \frac{\sqrt{\p}}{2} \, \frac{1}{x}
                + \CO \left( \frac{1}{x^2} \right) \qd, \\
     J_+(x) & = & \frac{1}{x} \,
                  \int_0^\infty dk \, k e^{- (k + k_0)^2/T}
                  + \CO \left( \frac{1}{x^2} \right) \qd.
\eea
Hence
\beq
     J_-(\La /z) = (z/\La) \left( k_0 \sqrt{\p T} +
                   \int_0^\infty dk \, k e^{- (k + k_0)^2/T} \right)
                   + \CO (z^2) \qd.
\eeq
We can further show that
\beq
     \lim_{x \rightarrow 0+} \left\{ J_+(x) - \tst{\2}
        \left(  \sqrt{- T \ln(x)} - k_0 \right)^2 \right\} = 0 \qd,
\eeq
which implies that for small positive $z$
\beq
     J_+(z) = - T \ln(z)/2 + \CO \left( \sqrt{- \ln(z)} \right) \qd.
\eeq
Then
\beq \label{flim}
     f(z) = \frac{- \ln(z)}{z} \, \frac{T \La}{2} \left(
            k_0 \sqrt{\p T} + \int_0^\infty dk \, k e^{- (k + k_0)^2/T}
	    \right)^{-1} + \CO \left( \frac{\sqrt{- \ln(z)}}{z} \right)
	    \qd.
\eeq
Eq.\ (\ref{flim}) shows that $\lim_{z \rightarrow 0+} f(z) = + \infty$.
Since $f(1) \le 1$, eq.\ (\ref{fofone}), we conclude that the saddle
point equation (\ref{saddleap}) always has a real positive solution
in the interval $0 < z \le 1$.

At low temperatures we can solve the saddle point equation
(\ref{saddleap}) ex\-plicit\-ly. Recall that we are dealing with the gas
phase $|B| + \m < 0$. In this phase
\beq
     \La^{-1} = (1 + e^{- 2|B|/T}) e^{(|B| + \m)/T}
\eeq
becomes a small parameter as $T$ goes to zero. We will solve the
saddle point equation (\ref{saddleap}) selfconsistently. Let us assume
that
\beq \label{zcons}
     \La^{-1} e^{- k_0^2/T} \; \ll \; |z| \; \ll \; \La \qd.
\eeq
We further assume that $k_0 > 0$, since the case $k_0 = 0$ can be
treated for all temperatures (cf.\ eq.\ (\ref{fofone})). Define
\beq
     I_\pm = \int_0^\infty dk \, k e^{- (k \pm k_0)^2/T} \qd.
\eeq
(\ref{zcons}) implies that $|\La/z| \gg 1$, and thus to leading order
\beq \label{jmlow}
     J_- (\La/z) = \frac{z I_-}{\La} \qd.
\eeq
On the other hand $|\La z| e^{k_0^2/T} \gg 1$, which implies that
\beq \label{jplow}
     J_+ (\La z) = \frac{I_+}{\La z} \qd.
\eeq
Inserting (\ref{jmlow}) and (\ref{jplow}) into the saddle point equation
(\ref{saddleap2}) we obtain two solutions $z_\pm = \pm z_c$,
\beq \label{zc1}
     z_c = \sqrt{I_+/I_-} \qd.
\eeq
The low temperature asymptotics of $I_+$ and $I_-$ are easily
calculated. We find
\bea
     I_+ & = & \left( \frac{T}{2 k_0} \right)^2 \, e^{- k_0^2/T}
                  \qd, \\ \label{imasy}
     I_- & = & k_0 \sqrt{\p T} + I_+ \approx k_0 \sqrt{\p T} \qd.
\eea
It follows that
\beq \label{zc2}
     z_c = \frac{T^{3/4}}{2 \p^{1/4} k_0^{3/2}} \, e^{- k_0^2/2T} \qd.
\eeq
Note that this solution is consistent with (\ref{zcons}).

In order to determine the directions of steepest descent let us
calculate $S''(z_\pm)$. First we rewrite $S(z)$, eq.\
(\ref{phaseap}), as
\beq
     S (z) = \frac{1}{\p} \int_0^\infty dk \, k
                   \ln \left\{
		   \frac{1 + (z \La)^{-1} e^{- (k + k_0)^2/T}}
		   {1 + \La^{-1} e^{- (k + k_0)^2/T}} \;
		   \frac{1 + z \La^{-1} e^{- (k - k_0)^2/T}}
		   {1 + \La^{-1} e^{- (k - k_0)^2/T}}
	           \right\} \, .
\eeq
For all $z$, which satisfy (\ref{zcons}), we obtain the low
temperature expansion
\bea
     S(z) & = & \frac{1}{\p \La} \left\{ \left( \frac{1}{z} - 1
                  \right) I_+ + (z - 1) I_- \right\} \nn \\
            & = & - 2 k_0 D \left\{ \left( 1 - \frac{1}{z} \right)
	            z_c^2 + 1 - z \right\} \qd, \label{phasyap}
\eea
where $D$ is the density, eq.\ (\ref{lowtdens}). Thus
\beq
     S''(z_\pm) = \pm \frac{4 k_0 D}{z_c} \qd,
\eeq
which means that $S''(z_+) > 0$, $S''(z_-) < 0$. Therefore the path
of steepest descent through $z_-$ is part of the real axis, whereas the
path of steepest descent through $z_+$ is perpendicular to the real
axis. The directions of steepest descent are depicted in figure 1. We
conclude from the figure that the relevant saddle point for the
calculation of our integral (\ref{leadint}) is $z_+$.
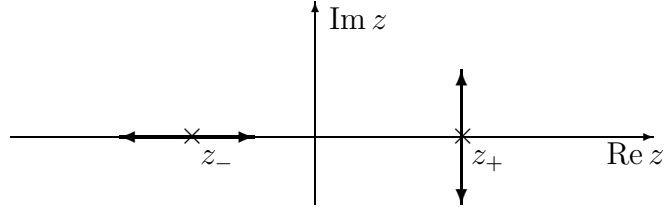
\begin{figure}

\unitlength .90mm

\begin{picture}(140,40)

\put(25,15){\vector(1,0){95}}
\put(70,5){\vector(0,1){30}}

\put(113,11){$\Re \, z$}
\put(72,31){$\Im \, z$}

\put(90,14){$\times$}
\put(50,14){$\times$}

\put(93,11){$z_+$}
\put(53,11){$z_-$}

\thicklines

\put(91.6,15){\vector(0,1){10}}
\put(91.6,15){\vector(0,-1){10}}
\put(51,15){\vector(1,0){10}}
\put(51,15){\vector(-1,0){10}}

\end{picture}

\caption{Directions of steepest descent}
\end{figure}

Now the saddle point contribution to the integral (\ref{leadint})
follows from eq.\ (\ref{phasyap}). We obtain
\beq \label{phsaddle}
     S(z_+) = - 2 k_0 D (z_+ - 1)^2 \approx -2 k_0 D \qd.
\eeq
The exponent $\nu(z)$, which determines the power law behavior of the
correlation functions, has the following low temperature expansion,
\beq \label{nulow}
     \nu(z) = - \; \tst{\frac{1}{2\p}} \ln(1 - (2 - z - z^{-1})
                \La^{-1} e^{- k_0^2/T}) \qd.
\eeq
Since $z_c$ satisfies (\ref{zcons}) and $z_c \ll 1$, we obtain
\beq \label{nusaddle}
     \nu(z_+) = - \; \frac{e^{- k_0^2/T}}{2\p z_c \La} =
        - \; \frac{2D k_0^{3/2} e^{- k_0^2/2T}}{\p^{1/4} T^{5/4}} \qd.
\eeq
Eqs.\ (\ref{phsaddle}) and (\ref{nusaddle}) determine the saddle point
contribution to $\Om_\pm$,
\beq \label{omsaddle}
     \Om_\pm \sim t^{- (1 \pm \i \nu(z_+))} \, e^{- xD} \qd.
\eeq
Here we suppressed $\nu^2(z_+)$, which is small compared to 1.

The function $F(z)$, eq.\ (\ref{capitalf}),  has two simple poles at
$z = \g^{\pm 1}$, which may contribute to the integral (\ref{leadint}),
if they are crossed in the process of deformation of the path of
integration. Let us investigate this question in the low temperature
limit. The cases $B > 0$, $B = 0$ and $B < 0$ have to be studied
separately.
\begin{enumerate}
\item
$B > 0$: In this case $\g = e^{2B/T}$, $\g^{-1} = e^{- 2B/T}$.
Since $z_+$ becomes small at low temperatures, we may have crossed
the pole at $\g^{-1}$. The condition for having no contribution from
the pole is $e^{- 2B/T} < z_+$, or, using (\ref{zc2}), $k_0^2 < 4B$.
Let us assume that
\beq \label{crosscond}
     k_0^2 > 4B \qd.
\eeq
Then $z = \g^{-1}$ satisfies (\ref{zcons}), and we can use the low
temperature expansion (\ref{phasyap}) to obtain the contribution of
the pole,
\bea
     S(\g^{-1}) & = & - \; 2 k_0 D ((1 - \g) z_c^2 + 1 - \g^{-1})
                          \nn \\
                  & \approx & - \; 2 k_0 D (1 - \g z_c^2)
		    \approx - \; 2 k_0 D \qd.
\eea
Here we used (\ref{crosscond}) once more to estimate that $\g z_c^2
\ll 1$. We see that the contribution of the pole to the exponential
decay is exactly the same as the contribution of the saddle point.
This is no longer true for the power law. (\ref{crosscond}) implies
that $\g \La^{-1} e^{- k_0^2/T} \ll 1$. Using this fact and the low
temperature expansion (\ref{nulow}) we obtain
\beq
     \nu(\g^{-1}) = - \; \frac{e^{(3B + \m - k_0^2)/T}}{2\p} \qd.
\eeq
\item
$B = 0$: In this case $\g = 2$, $\g^{-1} = \2$ and $z_+ < \g^{-1}$.
The pole at $\g^{-1}$ contributes to the integral (\ref{leadint}).
$z = \g^{-1} = \2$ satisfies (\ref{zcons}). Thus
\beq
     S(\g^{-1}) = - \; k_0 D (1 - 2 z_c^2) \approx - \; k_0 D \qd.
\eeq
Comparing this expression with the saddle point contribution
(\ref{phsaddle}) we see that the pole now yields the leading
contribution to the exponential decay. For $\nu(\g^{-1})$ we find
\beq
     \nu(\g^{-1}) = - \; \frac{e^{(\m - k_0^2)/T}}{2\p} \qd.
\eeq
\item
$B < 0$: Now both poles are very close to 1, $\g = 1 + e^{2B/T}$,
$\g^{-1} = 1 - e^{2B/T} + e^{4B/T}$. The pole at $\g^{-1}$
contributes to the integral, and $z = \g^{-1}$ satisfies
(\ref{zcons}). Using again (\ref{phasyap}) we obtain
\beq
     S(\g^{-1}) = - \; 2 k_0 D e^{2B/T} (1 - z_c^2)
		  \approx - 2k_0 D e^{2B/T} \qd.
\eeq
Since $B < 0$, this gives again the leading exponentially decaying
contribution to $\Om_\pm$. $\nu(\g^{-1})$ is obtained from
(\ref{nulow}),
\beq
     \nu(\g^{-1}) = - \; \tst{\frac{1}{2\p}}
        \ln(1 + e^{4B/T} \La^{-1} e^{- k_0^2/T})
        \approx - \; \frac{e^{(3B + \m - k_0^2)/T}}{2\p} \qd.
\eeq
\end{enumerate}
Using the explicit formulae for the density $D$, eq.\ (\ref{lowtdens}),
we obtain a uniform description of the cases (a), (b) and (c),
\bea \label{omnipadre}
     \Om_\pm & = & \oint dz \, F(z) \, t^{- \2(1 \pm \i \nu(z))^2}
                   e^{t S(z)} \nn \\
             & \sim & t^{- \left(\2 \pm \i \nu(\g^{-1})\right)} \,
	           \tst{\exp \left\{- \frac{x}{2} \sqrt{\frac{T}{\p}}
                   e^{(\m + B)/T} \right\}} \qd, \qd\\
     \nu(\g^{-1}) & = & - \; \frac{e^{(3B + \m - k_0^2)/T}}{2\p} \qd.
\eea
For a physical interpretation of eq.\ (\ref{omnipadre}) we look at the
formula (\ref{densab}) for the density of down-spin electrons. Its
low temperature expansion is
\beq \label{lowdown}
     D_\ab = \tst{\2 \sqrt{\frac{T}{\p}} e^{(\m + B)/T}} \qd.
\eeq
Inserting this expression into (\ref{omnipadre}) we arrive at eq.\
(\ref{omum}) of the main text. Furthermore, comparing (\ref{lowdown})
and (\ref{lowtdens}) we see that we can identify $D_\ab$ with $D$, when 
the magnetic field is positive.Hence we may replace $D$ by $D_\ab$ in
eqs.\ (\ref{nusaddle}) and (\ref{omsaddle}), which implies eqs.\
(\ref{edens}) and (\ref{nuzp}) in section~7.

We still have to discuss the range of validity of the above low
temperature calculations. The low temperature expression for the saddle
point $z_c$ is sufficiently precise, as long as $z_c$ satisfies
(\ref{zcons}). Yet, there is another condition to be satisfied. Note
that the phase $S(z)$ in low temperature approximation, eq.\
(\ref{phasyap}), is multiplied by a factor which vanishes as the
temperature approaches zero. We have
\beq
     t S(z) = - 2 k_0 Dt \left\{ \left( 1 - \frac{1}{z} \right)
	            z_c^2 + 1 - z \right\} \qd.
\eeq
Thus the large parameter, which we need for the saddle point
approximation to be valid, is $xD = 2k_0Dt$ rather than $t$.


\end{document}